# Pathways of structural and magnetic transition in ferromagnetic shape memory alloys

Matthew R. Sullivan, Ashish A. Shah, and Harsh Deep Chopra*

*Thin Films and Nanosynthesis Laboratory, Materials Program, Mechanical and Aerospace Engineering Department, SUNY-Buffalo, Buffalo, NY 14260, USA*

## Abstract

A fundamental question in the study of ferromagnetic shape memory alloys (FSMAs) is: What is the nature of the magneto-elastic coupling in these alloys and to what extent does it drive structural transformation? This question also holds the key to developing new and optimized alloys that combine high strains at low switching field. In the present study, it is shown that the reconfiguration of the micromagnetic structure is enslaved to and follows the martensitic transformation in these alloys, using Ni-Mn-Ga and Fe-Pd systems. This is determined by developing a new high speed, electronic method to study temperature dependent domain dynamics, called 'magnetic transition spectra'. The sequence of structural and magnetic transitions was found to be as follows: Cooling: structural transition followed by micromagnetic reconfiguration; Heating: micromagnetic reconfiguration followed by structural transition.

## I. INTRODUCTION

Shape memory alloys (SMAs) provide large displacements and forces within a small actuator design. However, these alloys have slow dynamic response due to temperature controlled actuation. In recent years, an alternative approach to (faster) actuation has emerged through the use of SMAs that are ferromagnetic in nature.[1,2] Ferromagnetic shape memory alloys (FSMAs) offer an ability to cause actuation by an applied magnetic field rather than the slow process of shape change by temperature, thereby combining high strains with fast reaction times. The FSMAs are complex correlated systems whose physical properties are governed by interactions across different energy regimes (thermal, magnetic, and elastic). These interactions produce a rich variety of phenomena, such as, for example, the magneto-elastic shape memory effect,[1,2] thermo-elastic shape memory effect,[3] and the magneto-caloric effect.[4] Examples of FSMAs include the Fe-Pd, Fe-Pt, Fe-Ni, Fe-Ni-Co-Ti systems, and the family of Heusler alloys such as the $Ni_{2+x}Mn_{1-x}Ga$ system, which undergoes a reversible, thermoelastic martensitic phase transformation.

The ferromagnetic Weiss domains in FSMAs are magneto-elastically coupled to and superimposed on the martensite structure. Due to this coupling, when the geometrical configuration of the magnetic domains is altered by an applied magnetic field, it also leads to a change in the relative volume fraction of the martensite twins,[5,6,7] thus enabling field induced shape change. Practical applications require large strains at low switching fields. The magnitude of the switching field, in turn, depends on the magneto-elastic coupling, which governs the magnitude of the applied field either to cause a structural transformation or change the relative volume fraction of martensite twins. It is widely recognized that the trial and error method of designing new alloys would give way to a more scientific approach if the transformation pathways of structural and magnetic transitions can be understood. Whereas the structural transition is the well known displacive, diffusionless martensite transformation, the magnetic transition manifests itself as a redistribution of the magnetocrystalline anisotropy axis (both in strength and direction) from the austenite to the martensite phase. Therefore, a fundamental question arises: what is the nature of the magneto-elastic coupling in these alloys and to what extent does it drive structural transformation from the high temperature, higher symmetry austenite phase to the low temperature, lower symmetry martensite phase? In the present study, through the development of a simple technique called ‘magnetic transition spectra’ (MTS) to study temperature dependent domain dynamics, as well as other experimental results, it is shown that the evolution of the micromagnetic structure is enslaved to and follows the structural transformation in these alloys. The MTS method is also simple and versatile, and can be used to study other magnetic transitions, such as the onset of exchange anisotropy in ferromagnetic-antiferromagnetic systems, phenomenon of brittle-to-ductile transition in structural materials, and pressure induced phase transitions.

## II. EXPERIMENTAL DETAILS

Different off-stoichiometric {100} $Ni_{2+x}Mn_{1-x}Ga$ Heusler alloy single crystals as well as Fe-30at%Pd single crystals were investigated. The single crystals were grown by the modified Bridgeman method and crystal orientation was determined by the Laue back-reflection method. Prior to the experiments, the samples were mechanically polished, followed by chemical etching to remove the strained layer caused by polishing. The transformation temperatures were measured using a differential scanning calorimeter.

The temperature dependent evolution of the micromagnetic structure was studied using the high resolution Interference-Contrast-Colloid (ICC) technique. The ICC method[8] and micromagnetics of FSMAs are discussed in detail elsewhere.[5,6,7] Briefly, the ICC method employs a colloidal solution to decorate the microfield on a magnetic surface, similar to the versatile Bitter method.[9] However, the technique differs in the manner in which the colloid decorated microfield is detected. In the Bitter method, a problem in contrasts develops in the bright field or the dark field mode due to backscattering by particles and the various surfaces between the objective lens and the specimen, which results in an overall loss of resolution. Instead, the ICC method uses a Nomarski interferometer to detect the surface microfield distribution. The magnetic microfield on a magnetic surface causes local variation in the density of colloid particles (average colloid particle size is 7 nm), thereby delineating the domain structure. This microfield is detected by polarization interferometer optics, which detects any unevenness at the nanometer scale and reveals domain structure with a pronounced three dimensional effect and at a high resolution limited only by that of the microscope. In this manner, the micromagnetic structure is directly observed superimposed on the microstructure of the sample as a function of temperature. In order to prevent freezing of the (water-based) ferrofluids, an oil-based ferrofluid was used, which remains stable from 70 $^{o}$C to -35 $^{o}$C. Observations were made by placing the samples in a commercially available precision heating and cooling stage (from Linkam, England) with an accuracy of $\pm 0.1$ degrees.

**The technique of magnetic transition spectra:**

The structural transformation in FSMAs is martensitic, which owing to its displacive and diffusionless nature proceeds at a speed approaching the transmission of elastic waves in a solid, modified (and slowed) by an appropriate damping factor and softening of the elastic constants in the premartensitic state.[10-12] Accompanying the martensite transformation in FSMAs is a reconfiguration of the micromagnetic structure. Therefore a suitable method is required to provide simultaneous real-time information on both transitions. Whereas the martensitic transformation can be observed directly in an optical microscope using, say, a Nomarski interferometer, simultaneous monitoring of the high speed magnetic transition using established techniques (such as the polar Kerr effect, magnetic force microscopy, Bitter method, or the ICC method) proved to be too cumbersome and/or slow or inconclusive. This is especially true when the two different transitions occur at different speeds and have to be followed simultaneously as a function of temperature in a sample contained in the protective environment of a heating/cooling stage. Therefore a novel and simple method was developed, which we refer to as the 'magnetic transition spectra' (MTS) method.

The MTS method is an electronic method to monitor the change in micromagnetic structure as a function of temperature.[7,13] It is based on the same principle upon which the well known and established Barkhausen method is based,[14] namely, Faraday's law: voltage $\xi$ induced in a pickup coil is proportional to the rate of change of flux with time $\xi = -d\Phi/dt$ in a sample; $d\Phi$ is the flux change over a time interval $dt$. Whereas the Barkhausen method generates a spectrum of voltage spikes by placing a pickup coil next to a ferromagnetic sample and cycling the sample in an applied magnetic field, see for example, Ref. [15], the MTS method generates a spectrum of voltage by *sweeping the sample across the transformation temperatures* instead of applying any magnetic field. As a result of phase transformation, either during forward or reverse martensitic transition, change in flux occurs in the sample. The resulting spectrum of voltage spikes as a function of temperature is recorded by an adjacent pickup coil. The MTS method is especially well suited for studying magnetic transitions in FSMAs because the local change in flux occurs over short intervals of time, which in turn leads to a high output signal.

The experimental setup used in the MTS method is the same as that for the Barkhausen method.[15] However, unlike the Barkhausen method where an energizing coil is used to generate a magnetic field to sweep the sample, a heating/cooling stage is used in the MTS method, as shown schematically in Fig. 1. The MTS signal pickup coil is a specially designed and optimized miniature surface probe fabricated out of an insulating ferrite core. Its output is bandwidth limited by a band pass filter to a range of 3-10 kHz. The pickup coil is a miniature probe in the form of a hemispherical ring 2-3 mm in diameter, 1 mm thick and having 50 encircling coil turns. The sample is attached to the probe and placed in a commercial temperature stage (Linkam, England) that is capable of heating and cooling the sample from 77 K to 873 K to within $\pm$ 0.1 $^{o}$C. The heating/cooling stage is fully automated and interfaced with an image frame grabber (Linkam) and a synchronized high speed data acquisition card. Furthermore, the experimental setup acquires and labels all images and data automatically following the execution of a pre-programmed heating/cooling cycle. The software operating the stage and acquisition cards also has the ability to embed important experimental information (temperature, heating/cooling rate, etc.) directly on the recorded optical micrographs, which themselves are collated and numbered automatically by the software. This degree of automation ensures unambiguous correlation between the microstructure and the acquired MTS spectrum as a function of temperature. The samples were cooled/heated from rates as low as 0.1 $^{o}$C/min to 30 $^{o}$C/min. The MTS spectra were found to be independent of the measured heating/cooling rates because the transformation occurs at speeds much faster than the temperature scans. Moreover, the transition temperatures were found to be unaffected by the presence of the miniature pickup coil lying adjacent to the samples. As the sample is cooled or heated across its critical transformation temperatures, the signal from the MTS coil is digitized by the computer and analyzed subsequently.

## III. RESULTS AND DISCUSSION

Figure 2 shows the DSC curves of a Ni-Mn-Ga sample, which delineates the critical transformation temperatures for both forward and reverse martensitic transition. As seen from Fig. 2, and also confirmed by optical observations, the martensite start $M_s$ and finish $M_f$ temperatures are -19.3 $^oC$ and -19.6 $^oC$, respectively, and the austenite start $A_s$ and finish $A_f$ temperatures are -13 $^oC$ and -12.7 $^oC$, respectively; the sample had a Curie temperature of -103 $^oC$. Detailed temperature dependent micromagnetic studies show that the high temperature FCC phase exists in a large single magnetic domain state spanning several thousand microns, and this was a common feature for both Ni-Mn-Ga and Fe-Pd alloys.[6,7] The martensite transition results in the transformation of the structurally and magnetically homogeneous high temperature austenite phase into the heterogeneous low temperature martensite phase. Structurally, the 'heterogeneity' refers to the formation of fine twins, which are differently oriented variants of the lower symmetry martensite phase. Magnetically, whereas the FCC phase has a spatially well defined magnetocrystalline anisotropy axes throughout the volume of the crystal, the direction of the magnetocrystalline anisotropy in the martensite phase varies from one twin plate to another. This causes a micromagnetic reconfiguration from a single domain state in the FCC phase to a multiplicity of magnetic domains in the martensite phase. An example of the structurally and magnetically heterogeneous martensite state is shown in Fig. 3 for a Ni-Mn-Ga sample. In Fig. 3, the magnetic domains can be seen confined within the martensite bands that run diagonally across the micrographs from left to right. Also note the presence of 'inter-domain' walls that traverse numerous thick bands. As discussed in detail elsewhere, the domain structure in Fig. 3 is a result of a complex martensite structure, and can be explained by taking into account the fact that the thick martensite bands are not homogeneous entities but are in fact internally twinned.

Due to the high speed of the martensite transformation, a catastrophic or avalanche-like reconfiguration of the magnetic domain structure occurs in the crystal on passing from the single domain FCC phase to multi-domain FCT phase. This gives rise to a spectrum of voltage spikes reflecting the dynamics of the magnetic transition. Figure 4(a) shows the acquired magnetic transition spectrum accompanying the martensite transformation in the Ni-Mn-Ga sample whose DSC curve is shown in Fig. 2. As shown in Fig. 4(a) the MTS consists of a very large number of voltage spikes reflecting the dynamics of micromagnetic reconfiguration. Previously, we have used a 'jumpsum' method to analyze and interpret voltage spectra (Barkhausen spectra) in ferromagnetic materials.[15-17] The jumpsum analysis method is well suited because instead of assigning an average or mean value to a given spectrum, it expresses the acquired signal in terms of the profile of the spectrum. Thus with respect to Fig. 4(a), one of the signal parameters extracted from the MTS is called the JumpSum $JS$, which is simply the running total of all the voltage jump heights, as shown in Fig. 4(b) – the reverse S-curve. The $i^{th}$ value of the $JS$ is equal to the sum of all preceding voltage jumps $\sum_{i=1}^{i=i} \xi_i$ up to that temperature on cooling. The inset in Fig. 4(b) is a magnified view of a portion of the $JS$ curve, which shows that the $JS$ curve is in fact made up of a large number of jumps or steps. The $M_s$ and $M_f$ temperatures for this alloy were measured using both calorimetery and optical observations, and are also indicated in Fig. 4. As seen from Fig. 4(b), $JS$ value is zero above the $M_s$ temperature and rises to a saturation value within a narrow temperature interval $\Delta T = T_s - T_f$ of 1.5 degrees. In other words, *the structural transition precedes the micromagnetic reconfiguration on cooling*. Another parameter derived is called the JumpSum Rate $JSR$, which is simply the rate at which flux is being emitted by the sample. A composite of $JS$ and $JSR$ curves for both cooling and heating cycles for the Ni-Mn-Ga sample is shown in Fig. 5; similar curves were obtained for the Fe-Pd alloys. [Note that the $JS$ and $JSR$ curves in Fig. 4 and Fig. 5 start from right to left for cooling, and from left to right for heating]. The noteworthy feature of $JS$ and $JSR$ curves in Figs. 4 and 5 is the clear indication that during cooling, the magnetic transition occurs after the structural transformation, whereas on heating the magnetic transition is complete before the structural transition begins. In other words, the sequence of structural and magnetic transitions was found to be as follows: *Cooling: structural transition*$\rightarrow$*magnetic transition; Heating: magnetic transition*$\rightarrow$*structural transition*.

The precedence of the martensite transition over magnetic transition, say, during cooling, is further highlighted in Fig. 6, which shows a collage of micrographs that were acquired *in-situ* with the MTS measurements and overlaid on the $JS$ curve of Fe-30at%Pd alloy. Unlike the Ni-Mn-Ga system where the martensite transition occurs by the formation of very fine twins through the motion of a single interface across the surface of the sample, martensite transformation in the Fe-Pd alloy occurs by the abrupt, burst-like transformation of large volumes of the crystal into thick twin bands. This jerky and burst-like mode of transformation makes it easier to further illustrate the relative sequence of magnetic and structural transitions. The collage in Fig. 6 shows that *the formation of each set of twin bands leads to a sharp increase in the* $JS$ *value*.

<u>Note:</u> As a visual aid, the temperature dependent evolution of the micromagnetic structure with respect to the structural transformation in Fe-Pd alloy is shown in a movie that is submitted as a Supplementary Material along with this manuscript. The movie shows real time, temperature dependent evolution of the microstructure and the micromagnetic structure in a Fe-Pd alloy. The movie starts with the sample in the single domain FCC phase above the martensite temperature. On cooling, the reconfiguration of the magnetic domains can be clearly seen lagging behind the martensite transformation. Due to the abrupt martensite transformation, the magnetic structure finds itself incongruous with the new boundary condition imposed by the formation of the twin. The ensuing motion and self-accommodation of the magnetic domains into a micromagnetic structure that is consistent with the martensite structure is clearly evident in the movie. On heating, the magnetic domains can be seen moving and coalescing together well before the reverse martensite transformation begins.

Finally, the MTS method is simple to implement and can be used to study other magnetic phase transitions driven by *any* external influence that would cause an abrupt change in the micromagnetic state of the sample (for example, change in temperature, pressure, etc.). For example, we have successfully applied this method to study the onset of coupling in exchange anisotropy ferromagnetic films coupled to an antiferromagnetic substrate, using the Co-CoO systems.[18] Other potential applications of MTS include fundamental investigation of ductile to brittle transition in ferromagnetic structural materials and pressure-induced phase transitions. One limitation of the method, however, is the upper temperature limit. While it is easy to implement the technique for low temperature studies, it would need substantial modifications, both in the probe design and signal processing, when the temperatures exceed 70-80 $^{o}$C. Nonetheless, the MTS method could find applications in many fundamental investigations.

## IV. CONCLUSIONS

In conclusion, the present study shows that the micromagnetic reconfiguration during structural transformation in FSMAs is completely enslaved to the structural transformation. The sequence of structural and magnetic transitions was found to be as follows: Cooling: structural transition followed by micromagnetic reconfiguration; Heating: micromagnetic reconfiguration followed by structural transition. The MTS method is simple and can be adapted to study other phenomena where an abrupt change in flux occurs, for example, due to change in temperature or pressure.

## V. ACKNOWLEDGEMENTS

This work was supported by the DOE, Office of Basic Energy Science, Grant No. DE-FG02-01ER45906, and this support is gratefully acknowledged.

## REFERENCES

*Corresponding author. E-mail: hchopra@eng.buffalo.edu

## SUPPLEMENTARY INFORMATION

**The following Supplementary Information is submitted along with this manuscript:**

Movie showing the evolution of micromagnetic and structural transformation is the Fe-Pd sample. For ease, the optical contrast during the reverse martensite transformation (heating cycle) was changed slightly to distinguish it from the cooling cycles. See manuscript for discussion.

## FIGURE CAPTIONS

**FIGURE 1.** Schematic of the experimental setup for acquiring the magnetic transition spectrum. (T/C: thermocouple).

**FIGURE 2.** DSC curves for the forward (cooling) and reverse (heating) martensite transformation in a Ni-Mn-Ga sample.

(COLOR) **Figure 3.** Micrograph showing the magnetic domain structure of the FCT martensite in a Ni-Mn-Ga sample. Note the presence of domains within the bands as well as inter-domain walls spanning several bands.

**FIGURE 4.** (a) Magnetic transition spectrum for a Ni-Mn-Ga sample on cooling from a sample whose DSC curve is shown in Fig. 2. Note that the jumps occur below the martensite transformation. (b) The $JS$ curve corresponding to the MTS in (a). The inset in (b) is a magnified view show that the $JS$ curve consists of a large number of jumps.

(COLOR) **FIGURE 5.** Composite of $JS$ and $JSR$ curves for both cooling and heating. Note: For cooling the $JS$ curve goes from right to left whereas for heating the $JS$ curve goes from left to right.

(COLOR) **Figure 6.** Collage showing the $JS$ curve for Fe-30at%Pd sample and the corresponding change in microstructure responsible for each jump in the $JS$ curve.

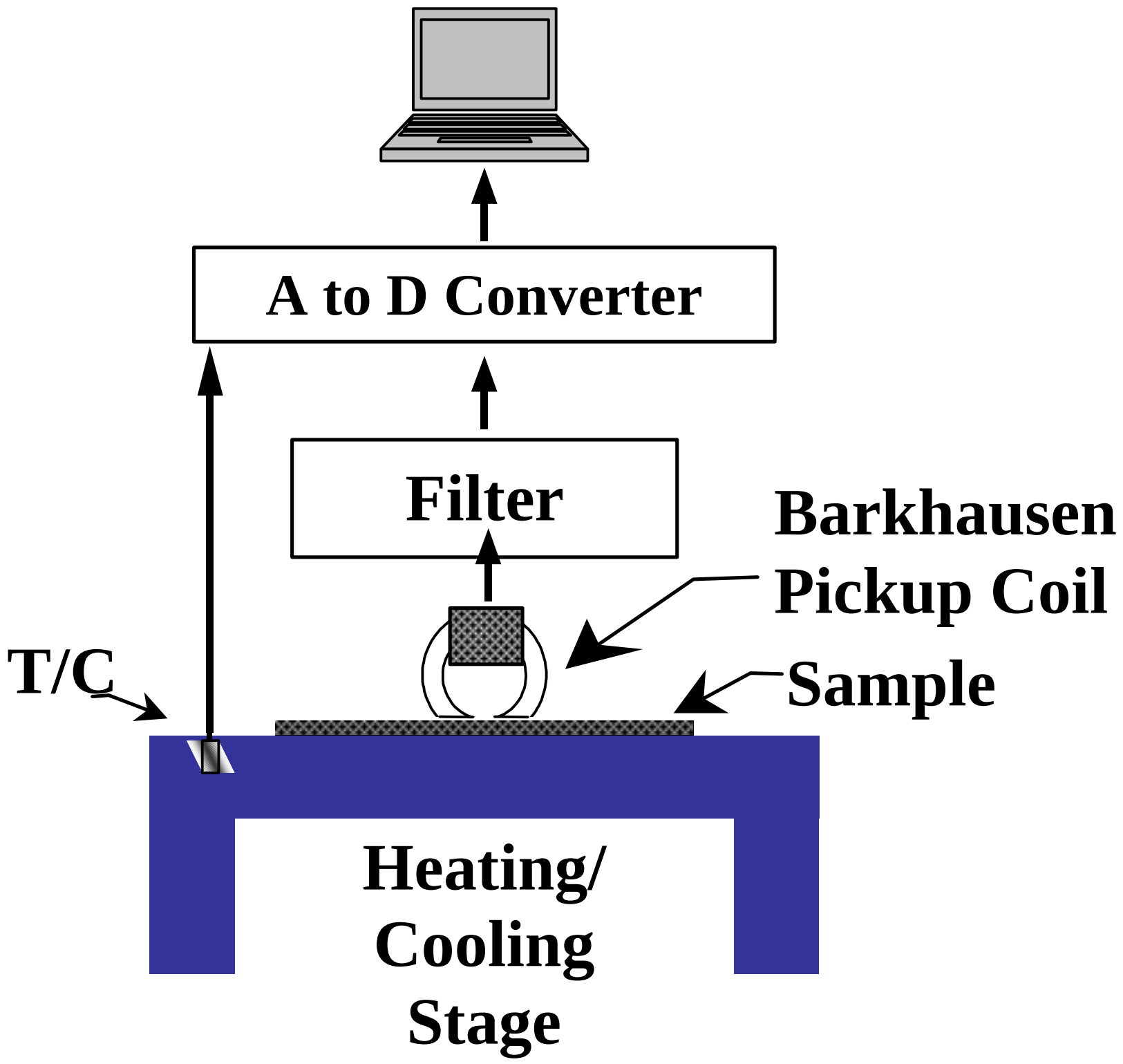

**Figure 1**

Chopra *et al.*, Phys. Rev. B

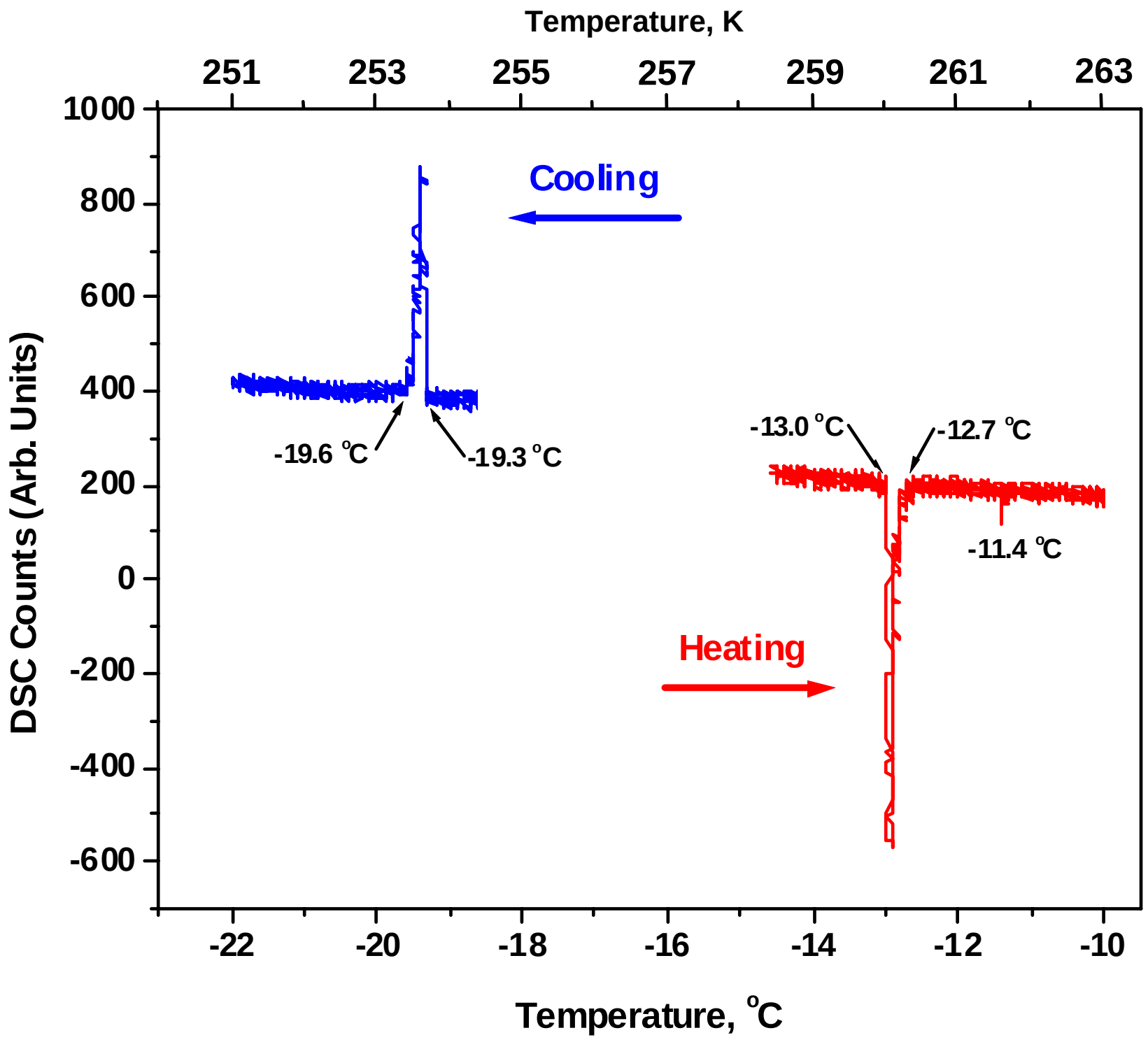

**Figure 2**

Chopra *et al*., Phys. Rev. B

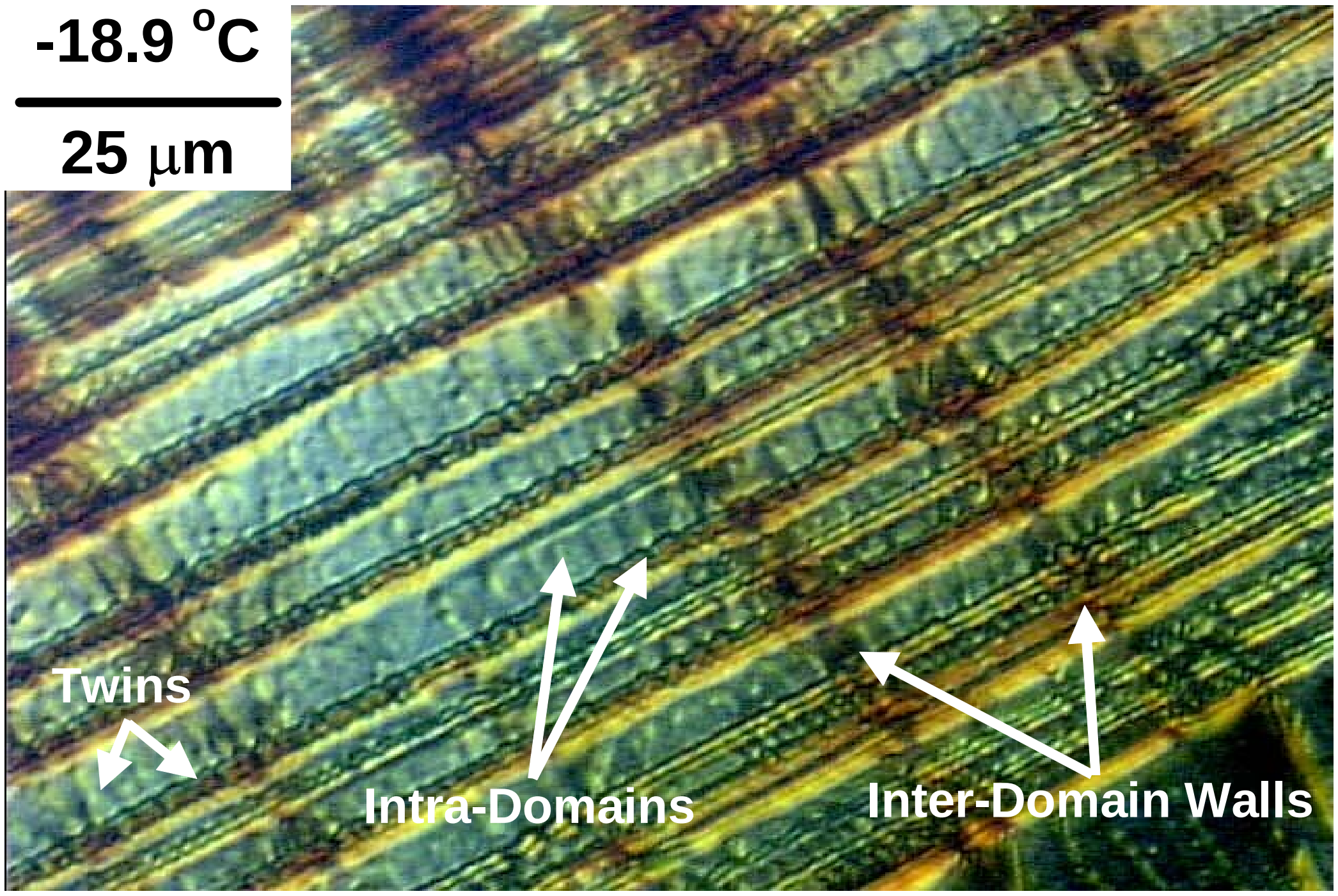

**Figure 3 (Color)**

Chopra *et al.*, Phys. Rev. B

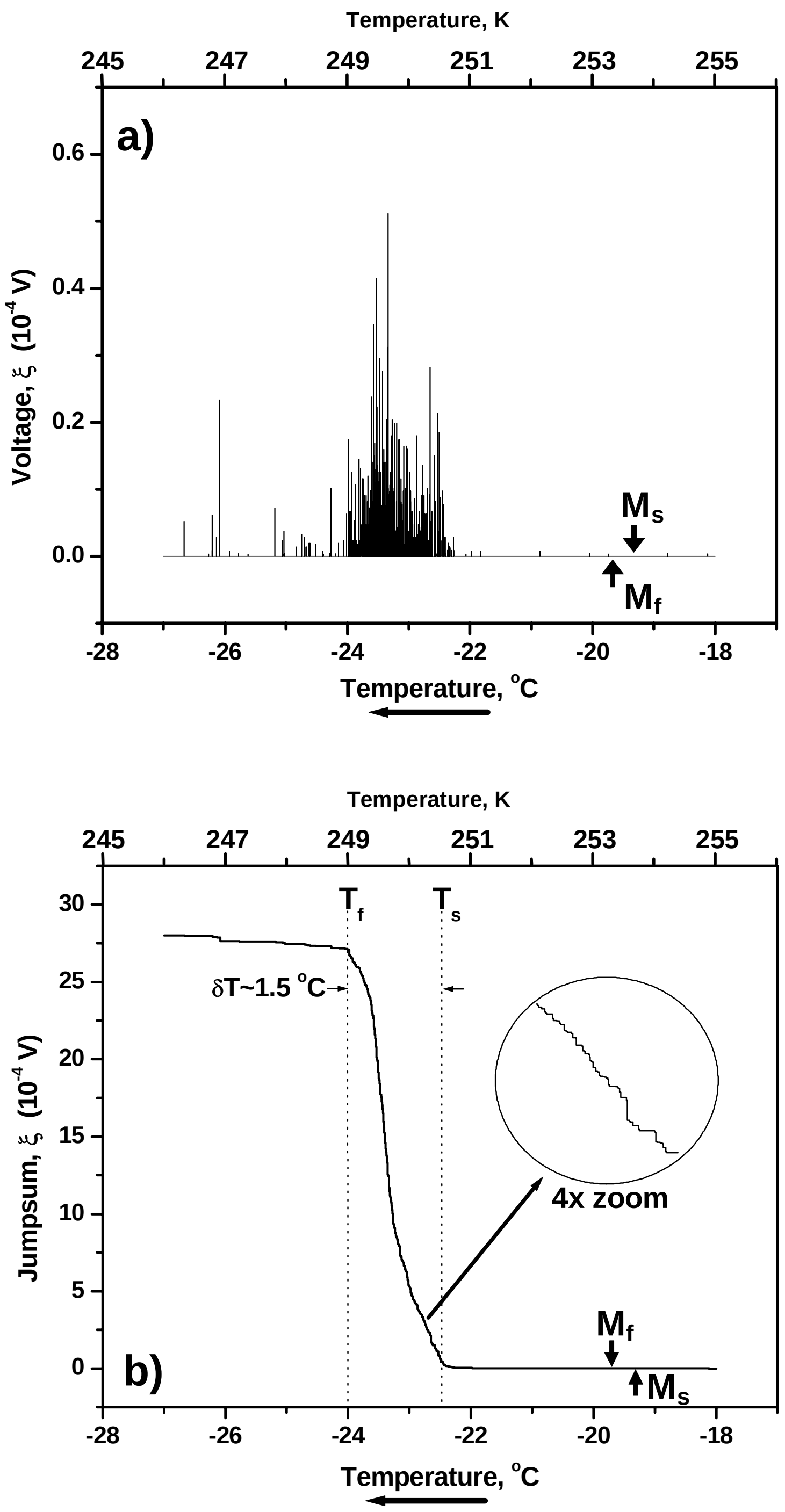

**Figure 4**

Chopra *et al.*, Phys. Rev. B

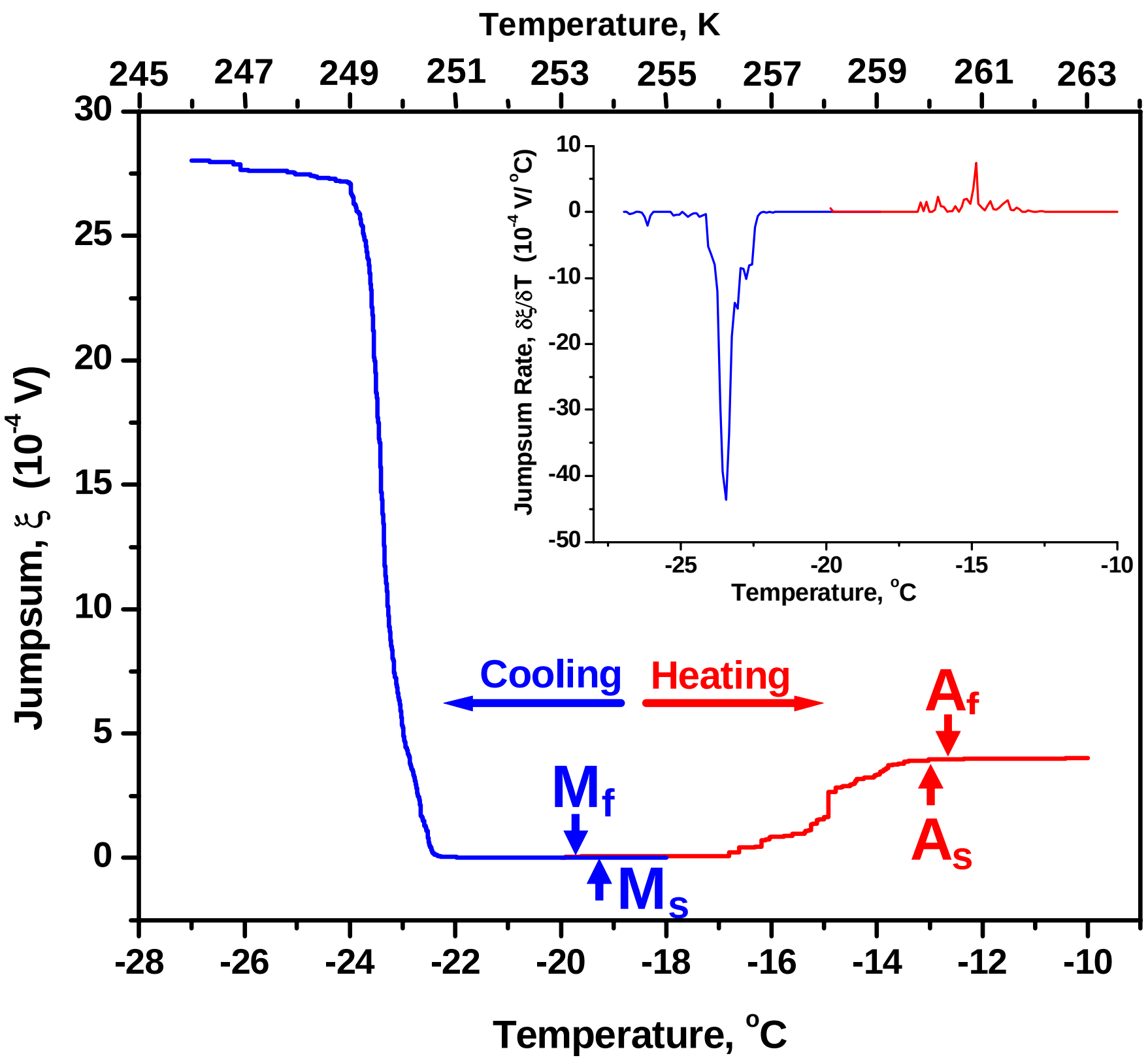

**Figure 5 (Color)**

Chopra *et al.*, Phys. Rev. B

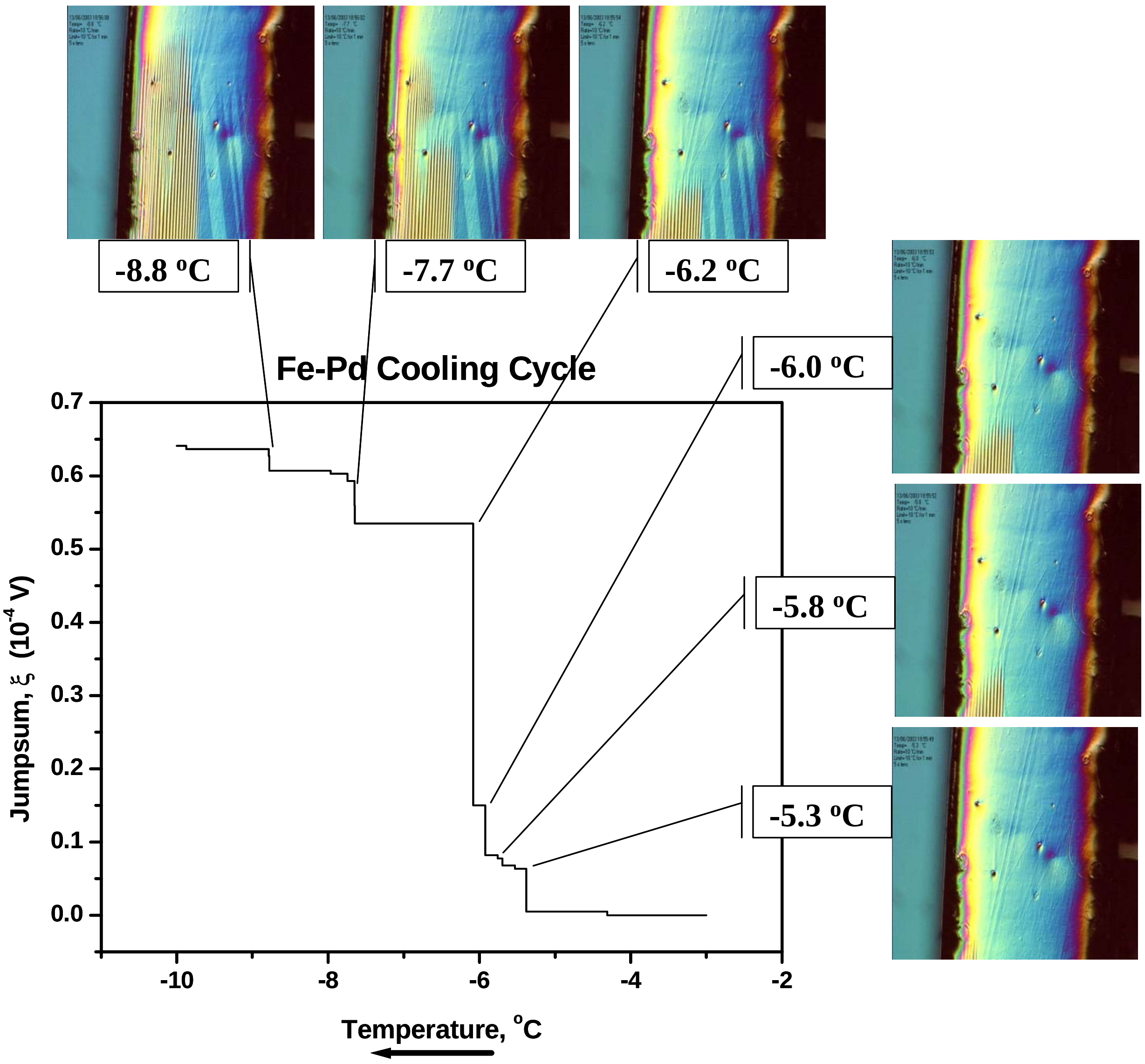

**Figure 6 (Color)**

Chopra *et al.*, Phys. Rev. B